\documentclass{pnastwo}
\usepackage{graphicx}
\usepackage{amssymb,amsfonts,amsmath}
\usepackage{url}

\begin{document}

\title{Multirelational organization of large-scale social networks in an online world}

\author{Michael Szell\affil{1}{Section for Science of Complex Systems, Medical University of Vienna, Spitalgasse 23, 1090 Vienna, Austria}, Renaud Lambiotte\affil{2}{Institute for Mathematical Sciences, Imperial College London,  London SW7 2PG, United Kingdom}\affil{3}{Department of Mathematics, Imperial College London,  London SW7 2AZ, United Kingdom} \and Stefan Thurner\thanks{To whom correspondence should be addressed. E-mail:
stefan.thurner@meduniwien.ac.at} \affil{1}{}\affil{4}{Santa Fe Institute, 1399 Hyde Park Road, Santa Fe, NM 87501, USA}}

\contributor{}

\maketitle

\begin{article}

\begin{abstract} The capacity to collect fingerprints of individuals in online media has revolutionized the way researchers explore human society. Social systems can be seen as a non-linear superposition of a multitude of complex social networks, where nodes represent individuals and links capture a variety of different  social relations. Much emphasis has been put on the network topology of social interactions, however, the multi-dimensional nature of these interactions has largely been ignored, mostly because of lack of data. Here, for the first time, we analyze a complete, multi-relational, large social network of a society consisting of the 300,000 odd players of a massive multiplayer online game.
We extract networks of six different types of one-to-one interactions between the players. Three of them carry a positive connotation (friendship, communication, trade), three a negative (enmity, armed aggression, punishment). We first analyze these types of networks as separate entities and find that negative interactions differ from positive interactions by their lower reciprocity, weaker clustering and fatter-tail degree distribution.
We then explore how the inter-dependence of different network types determines the organization of the social system. In particular we study correlations and  overlap between different types of links and demonstrate the tendency of individuals to play different roles in different networks. 
As a demonstration of the power of the approach we present the first empirical large-scale verification of the long-standing structural balance theory, by focusing on the specific  multiplex network of friendship and enmity relations. 
\end{abstract}


\dropcap{H}uman societies can be regarded as large numbers of locally interacting agents, connected by a broad range of social and economic relationships. These relational ties are highly diverse in nature and can represent e.g. the feeling a person has for another (friendship, enmity, love), communication, exchange of goods (trade) or behavioral interactions (cooperation or punishment). Each type of relation spans a social network of its own. 
A systemic understanding of a whole society can only be achieved by understanding these individual networks  and how they influence and coconstruct each other. The shape of one network influences the topologies of the others, as networks of one type may act as a constraint, an inhibitor, or a catalyst on networks of another type of relation.
For instance, the network of communications poses constraints on the network of friendships, trading networks are usually constrained to positively connoted interactions such as trust, and networks representing hostile actions may serve as a catalyst for the network of punishments. 
 A society is therefore characterized by the superposition of its constitutive socio-economic networks, all defined on the same set of nodes. This superposition is usually called multiplex, multi-relational, multi-modal or multivariate network, see Fig.~1. The study of small-scale multiplex networks has a long tradition in the social sciences \cite{wasserman}, and has been applied to areas such as homophily in social networks \cite{mcpherson}, the effect of combined interactions on an agent's behavior \cite{ajs} and the non-trivial inter-relation between family and business networks \cite{padgett}. Multiplexity is thought to play an important role in the organization of large-scale networks. For example the existence of different link types between agents explains the overlap of community structures observed in social networks, where nodes may belong to several communities, each associated to one different type of interaction \cite{EL09,ABL09}. Methodological work on multiplex networks includes the development of multiplex community detection \cite{multiplex}, clustering \cite{selee2007ecl} and other network analysis algorithms \cite{rodriguez2009emr}. The role of multiple relation types in measured social networks has recently been investigated across communication media \cite{baym2004sia}, in an online game \cite{huang2009vte}, as well as in ecological networks \cite{genini2010cmn}.

\begin{figure}[t]
\includegraphics[width=0.42\textwidth]{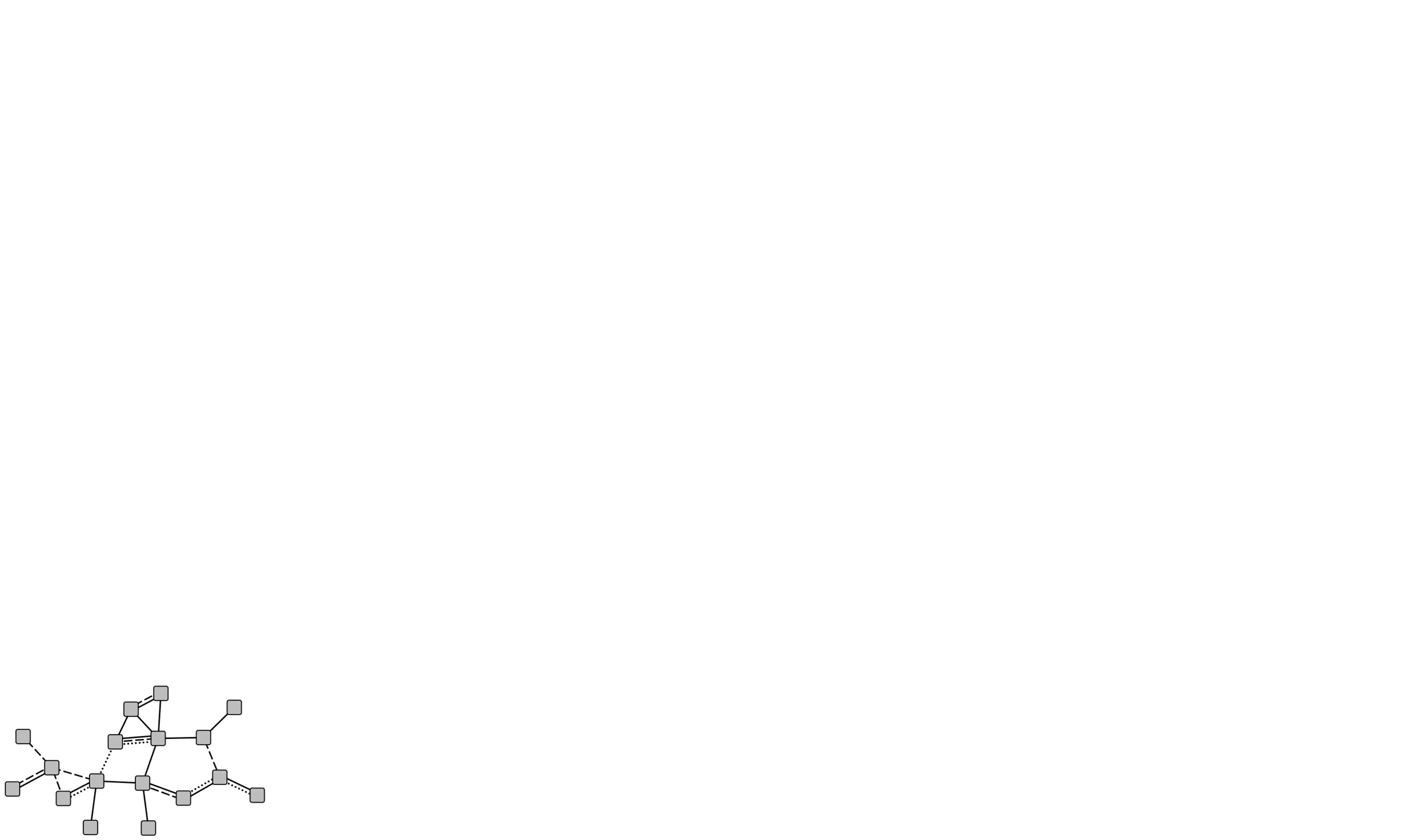}
\caption{Multiplex networks consist of a fixed set of nodes connected by different types of links. This multi-relational aspect is usually neglected in the analysis of large social networks. In our MMOG data-set, six types of social links can exist between any two players, representing their friendship or enmity relations, their exchanged private messages, their trading activity, their one-to-one aggressive acts against each other (attacks), and their placing of head-money (bounties) on other players as e.g. means of punishment.}\label{fig1}
\end{figure}

Traditional methods of social science, such as  small-scale questionnaire-based approaches, get more and more replaced by automated methods of data collection which allow for entirely different scales of analysis \cite{review,newmanreview, boccaletti2006cns}. This change of scale has opened new perspectives and has the potential to radically transform our understanding of social dynamics  and organization \cite{perspectives}. 
The empirical verification of social theories such as the strength of weak ties \cite{onnela,szell} become possible with hitherto unthinkable levels of precision. 
However, this large-scale perspective suffers from the drawback of a relatively coarse-grained representation of social processes taking place between individuals and of blindness in respect to the existence of different types of social interactions.
For example in most works on email \cite{email} or mobile phone networks \cite{onnela,mobile}, the existence and weight of a link is determined by the volume of information exchanged between two individuals. 
Although nodes can be generally well characterized (age, sex, zip code, etc.), the corresponding type of interaction (e.g. family or work interaction), is usually unavailable in the data and can only be inferred from behavioral patterns \cite{eagle}. Moreover, research on large social networks has focused on single types of interaction only, e.g. phone or email communication, and has ignored the wide spectrum of human interactions in real life \cite{mcpherson}. Whenever inter-dependencies and feedbacks between multiple relational interactions are significant, an aggregate representation of the different network types, or the representation of one single type will lead to a biased and misleading characterization of the organization of the system.

The following work is an attempt toward fully characterizing the multiplex nature of a large-scale social system. 
To this end we analyze coherent data from a complete society consisting of about 300,000 players of a massive multiplayer online game (MMOG) \cite{castronova2005swb}.
Having become extremely popular over the past years, there exists a multitude of large-scale online games -- often played by thousands, sometimes even millions. These games offer the possibility to experience alternative lives in which players can engage in different types of social interactions, ranging from establishing friendships and economic relations to the formation of groups, alliances, fighting and even waging of war \cite{szell}. Practically all actions of all players can be recorded in log files. The booming popularity of MMOGs opens previously unthinkable potentials for data-driven, quantitative socio-economic research \cite{virtual}, and enables e.g. economic surveys \cite{castronova2006orv}, studies on group dynamics \cite{mmogsimilar}, or large-scale social network analyses and the testing of classical sociological hypotheses \cite{szell}.

The data allows to identify the nature of one-to-one interactions between players; the topological properties of the corresponding networks -- defined on the same set of agents -- can be studied. 
We show that different types of interactions are characterized by distinct connectivity patterns. 
Exploring the inter-dependence of the different networks  reveals how multiplexity shapes the organization of the system at different levels, from the stability of local motifs to the global overlap between the networks. Moreover, the existence of positively  and negatively connoted interactions between players, e.g. through declared friendship or enmity, allows to analyze the organization of the system from the point of view of signed networks \cite{wasserman}.  Within this framework it becomes possible to experimentally verify structural balance \cite{bookKleinberg}, a long-standing theory in social psychology  \cite{heider1946aac} proposed for understanding emergence of conflict and tension in social systems \cite{cartwright}. The central idea behind structural balance is that some configurations of signed motifs, i.e. local `building blocks' of networks containing positive and/or negative ties, are socially and psychologically more stable than others and are therefore more likely to be present in human societies. By measuring the dynamics and abundance of signed triads (sets of three nodes connected by positive or negative links), we perform a large-scale validation of structural balance and provide insights indispensable for a realistic modeling of conflicts.

\begin{table*}[b]
\caption{
Single network properties. Properties of directed networks: number of nodes $N_{\alpha}$ (connected to at least one link), 
number of directed links $L_{\alpha}^{\mathrm{dir}}$, reciprocity $r_{\alpha}$ and in-degree/out-degree correlation 
$\rho(k_{\alpha}^{\mathrm{in}},k_{\alpha}^{\mathrm{out}})$. Greek indices mark network types. Properties of the corresponding undirected networks: 
number of undirected links $L_{\alpha}^{\mathrm{undir}}$, average degree $\bar{k}_{\alpha}$, clustering coefficient $C_{\alpha}$ and 
ratio to the corresponding random graph clustering $C_{\alpha}/C_{\alpha}^{\mathrm{rand}}$. The networks, when considered as separate entities, present distinct types of organization depending on the nature of the interactions. Positively (negatively) connoted links present high (low) values of $r_{\alpha}$, 
$\rho(k_{\alpha}^{\mathrm{in}},k_{\alpha}^{\mathrm{out}})$ and $C_{\alpha}$. 
}
\centering
\begin{tabular}{ ll|rrr|rrr|r}
			& 	& \multicolumn{3}{|c|}{Positive ties} & \multicolumn{3}{|c|}{Negative ties} & \\ \hline
			&				& Friends & PMs & Trades & Enemies & Attacks & Bounties & Envelope (all $\alpha$s) \\ \hline
directed 		&  $N_{\alpha}$  			& 4,313 & 5,877 & 18,589 & 2,906 & 7,992 & 2,980 & 18,819 \\
			& $L_{\alpha}^{\mathrm{dir}}$ 	& 31,929 & 185,908 & 796,733 & 21,183 & 57,479 & 5,096 & 967,205 \\ 
			& $r_{\alpha}$ 				& 0.68 & 0.84 & 0.57 & 0.11 & 0.13 & 0.20 & 0.59 \\ 
			& $\rho(k_{\alpha}^{\mathrm{in}},k_{\alpha}^{\mathrm{out}})$ & 0.88 &  0.98 & 0.93 & 0.11 & 0.64 & 0.31 & 0.95\\ \hline
undirected 	& $L_{\alpha}^{\mathrm{undir}}$ & 21,118 & 107,448 & 568,923 & 20,008 & 53,603 & 4,593 & 679,404 \\
			& $\bar{k}_{\alpha}$ & 9.79 & 36.57 & 61.21 & 13.77 & 13.41 & 3.08 & 72.20 \\
			& $C_{\alpha}$ & 0.25 & 0.28 & 0.43 & 0.03 & 0.06 & 0.01 & 0.42\\
			& $C_{\alpha}/C_{\alpha}^{\mathrm{rand}}$ & 109.52 & 45.71 & 131.95 & 6.13 & 37.27 & 13.88 & 109.93\\
\end{tabular}
\end{table*}

\section{Results}

\subsection{Nature of the Various Networks}
Different types of connectivity patterns may signal different organization principles behind the formation of networks \cite{newman,stan1}. 
Statistical properties of the six networks, when considered as separated entities, are collected in Table 1.  There,  
$N_{\alpha}$ is the number of nodes in the network $\alpha$, $L_{\alpha}^{\rm dir (undir)}$ is the number of (un)directed links. 
Reciprocity is labeled by $r_{\alpha}$, $\rho(k_{\alpha}^{\mathrm{in}},k_{\alpha}^{\mathrm{out}})$ is the correlation of in- and out degrees within the $\alpha$ network. Average degree, clustering coefficient, and clustering coefficient with respect to the corresponding random graph are marked by $\bar k_{\alpha}$, $C_{\alpha}$ and $C_{\alpha}/C_{\alpha}^{\mathrm{rand}}$, respectively. For definitions of the measures, see \cite{SI}. 

 {\em Positive links are highly reciprocal, negative links are not.}
Table 1 shows that networks with a positive connotation [friendship, private messages (PMs) and trades] are strongly reciprocal \cite{diego} (see \cite{SI}), in the sense that node pairs have a high tendency to form bi-directional connections, while networks with a negative connotation (enmity, attack and bounty) all show significantly smaller reciprocity. Low reciprocation in enemy networks may partially be explained by deliberate refusal of reciprocation to demonstrate aversion by total lack of response \cite{szell}. For attack networks, it may originate from the asymmetry in the strength of the players (a strong player is more likely to attack a weaker player to secure a win).
Asymmetry in negative relations is confirmed in the correlations between node in-degrees and out-degrees. Positive links are almost balanced in the in- and out-degrees, $\rho \sim 1$, whereas negative links show an obvious suppression in doing to others what they did to you. 

{\em Power-law degree distributions indicate aggressive actions.}
Studying cumulative in- and out-degree distributions, we find pronounced power-law distributions for aggressive behavior, i.e. attacking (out-degree for attacks), being declared an enemy (in-degree for enmity), and punishing/being punished (out- and in- degree for bounty). Power-laws are absent for positive (friendship, communication, trade) and passive links (being attacked), see Fig.~2. This discrepancy in degree distributions hints at qualitatively different link-growth/rewiring processes taking place in positive tie networks compared to the negative ones. For example, the classic network growth model of preferential attachment \cite{barabasi1999esr} leads to a power-law degree distribution. As we have shown in \cite{szell}, the growth of enemy networks is well characterized by this model, but not the growth of friend networks.

{\em Positive links cluster.}
From Table 1 it is clear that the positively connoted links show higher clustering coefficients than negatively connoted ones. High values of the clustering coefficient are expected for positive interactions due to their cohesive nature and the benefits of dense sub-graphs for better performance \cite{coleman}. The significantly lower values of clustering for negative values suggests that mechanisms such as triadic closure \cite{rapaport} are not dominant for negative interactions (see \cite{SI} for a confirmation), and has its origin in the balance of signed motifs (see below).

The independent analysis of the different networks reveals distinct types of organization which depend on the nature of the links. It is crucial to account for these distinct topological properties in models for the dynamics of cooperation and conflict in human societies. To demonstrate the danger of not differentiating between types of interactions we include data on the envelope network (as defined in \emph{Materials and Methods}) in Table 1. Neglecting the nature of social ties and mixing different interactions (even within the same data-set) results in gross mis-representation of the system, in this case at least by losing the typical low reciprocity and clustering observed in negative tie sub-networks.

For a detailed analysis of the time-evolution of single network properties on the same data set (first 445 days in the Artemis game universe), refer to \cite{szell}. There several `aging' or `maturing' effects were reported, such as a decrease of the clustering coefficient and reciprocity in friend networks over time.

\subsection{Network--Network Interactions}

Due to strong interactions between different social relations, a next level of complexity enters when considering the co-existence of different types of links \cite{raissa}. From now on, we only focus on undirected versions of the networks, as defined in \emph{Materials and Methods}. To quantify the resulting inter-dependencies between pairs of networks, we follow two approaches.

On one hand, we focus on the link-overlap between networks and calculate the Jaccard coefficient $J_{\alpha\beta}$ between two different sets of links $\alpha$ and $\beta$. The Jaccard coefficient quantifies the interaction between two networks by measuring the tendency that links simultaneously are present in both networks. 

\begin{figure}[t]
\hspace*{-0.15cm}\includegraphics{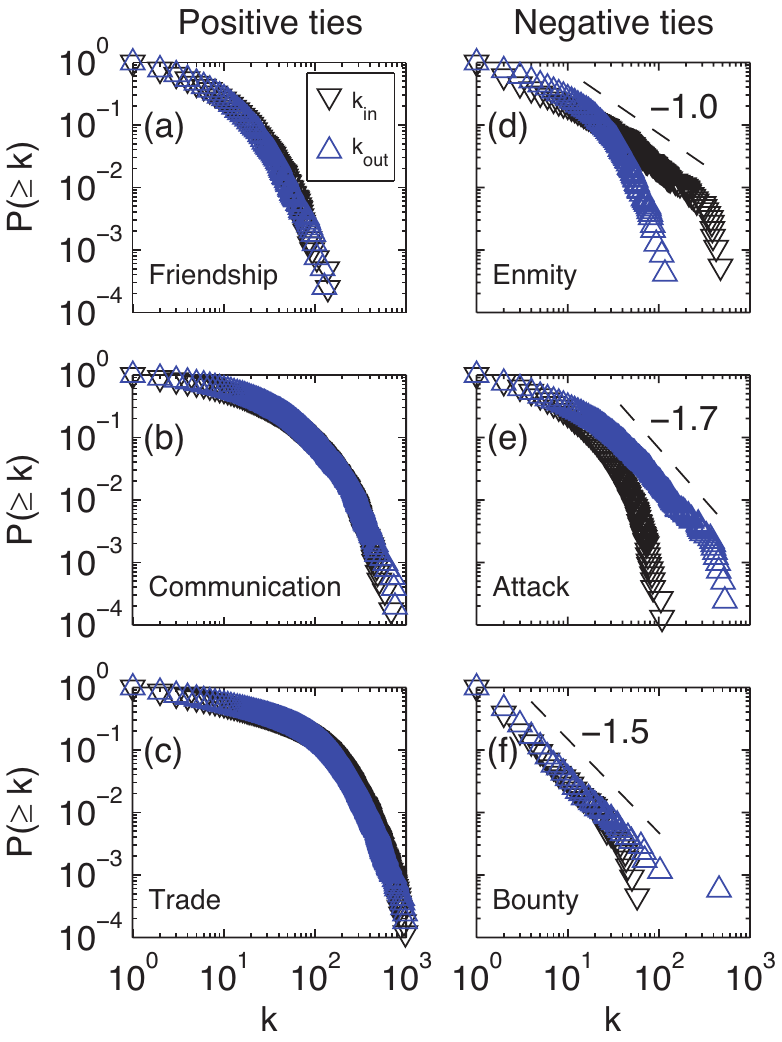}
\caption{Cumulative in-degree and out-degree distributions for the six types of networks spanning the same set of agents: (a) friendship, (b) communication, (c) trade, (d) enmity, (e) attack, (f) bounty. Note the differences between in- and out-degree distributions and the presence of power-laws (with cutoffs) for negatively connoted  interactions (right column), which are absent for positive ties (left column). It is immediately clear that topological properties of social networks depend strongly on the nature of their ties. Ignoring  this multi-relational composition can lead to loss of essential information.}\label{fig2}
\end{figure}

\begin{figure}[b]
\hspace*{-0.65cm}\includegraphics[width=0.49\textwidth]{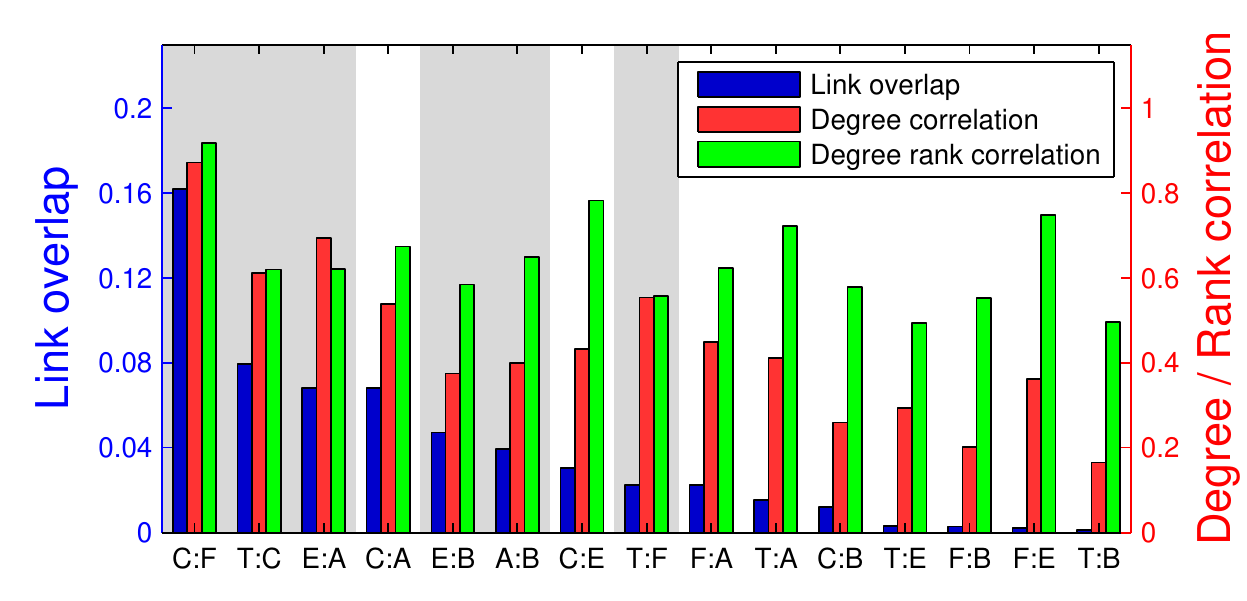}
\caption{Link overlap (Jaccard coefficient), degree correlation $\rho(k_{\alpha},k_{\beta})$ and degree rank correlation $\rho(\mathrm{rk}(k_{\alpha}),\mathrm{rk}(k_{\beta}))$ for all pairs of networks (ordered by link overlap), with the notations E for Enmity, F for Friendship, A for Attack, T for Trade, C for Communication and B for Bounty. Pairs of equal connotation (positive-positive or negative-negative) are marked with a gray background. These pairs have high overlaps, while oppositely-connotated pairs have lower overlaps. The various relations are organized in a non-trivial way, suggesting that agents play very different roles in different relational networks. 
 }\label{fig3}
\end{figure}

On the other hand, we compute correlations $\rho(k_{\alpha},k_{\beta})$ between node degrees in different networks (see \cite{SI}). These coefficients measure to which extent degrees of agents in one type of network correlate with degrees of the same agents in another one. If $\rho(k_{\alpha},k_{\beta})$ is close to $1$, players who have many (few) links in network $\alpha$ have many (few) links in network $\beta$. Note that both measures might be affected by different network sizes or average degrees. To account for this possibility, we additionally compute correlations $\rho(\mathrm{rk}(k_{\alpha}),\mathrm{rk}(k_{\beta}))$ between rankings of node degrees, where rk represents rank. Overlap and correlation quantities provide complementary insights into the organization of social structures. In Fig.~3 for all pairs of networks the three measures are shown. Note that no causal directions can be implied and that all correlations are positive. From highest to smallest overlap (from left to right), Fig.~3. provides the following conclusions:
\begin{itemize}
\item[] \emph{Communication--Friendship.} The pronounced overlap implies that friends tend to talk \emph{with each other}. The equally pronounced correlation attests that players who communicate with many (few) others tend to have many (few) friends. The former result was already reported in \cite{szell}, where a high fraction of communication partners was shown to be friends.
  \item[] \emph{Trade--Communication.} The high overlap shows that trade partners have a tendency to communicate with each other, while the high correlations shows a tendency of communicators being traders. 
  \item[] \emph{Enmity--Attack.} The high overlap shows that enemies tend to attack each other, or that attacks are likely to lead to enemy markings. The high correlations imply that aggressors or victims of aggression tend to be involved into many enemy relations.  
  \item[] \emph{Communication--Attack.} The relatively high overlap shows that there is a tendency for communication taking place between players who attack each other. The relatively high correlation implies that players who communicate with many (few) others tend to attack or be attacked by many (few) players.  Aggression is not anonymous, but accompanied by communication.
  \item[] \emph{Enmity--Bounty and Attack--Bounty.} Similar to Enmity--Attack.
  \item[] \emph{Communication--Enmity.} Similar to Com\-mu\-ni\-ca\-tion--Attack. 
  \item[] \emph{Trade--Friendship.} Similar to Trade--Communication, however with a smaller overlap. It is more difficult for traders to become friends than to just communicate.
  \item[] \emph{Friendship--Attack.} The low overlap shows that attacks tend to \emph{not} take place between friends, or that fighting players do \emph{not} tend to become friends. The relatively high correlations mean that players with many (few) friends attack or are attacked by many (few) others.
  \item[] \emph{Trade--Attack.} Similar to Friendship--Attack.
  \item[] \emph{Communication--Bounty.} Similar to Communication--Attack and Communication--Enmity, however with much smaller overlap and degree correlations.
  \item[] \emph{Trade--Enmity.} For this and all other interactions, overlap vanishes. Players who trade with each other almost never become enemies and vice versa. 
  \item[] \emph{Friendship--Bounty.} Similar to Com\-mu\-ni\-ca\-tion--Bounty.
  \item[] \emph{Friendship--Enmity.} The degree (rank) correlation is substantial, suggesting that players who are socially active tend to establish both positive as well as negative links. However, the vanishing overlap shows the absence of ambivalent relations. Friends are never enemies.
	\item[] \emph{Trade--Bounty.} This interaction shows the smallest values for all three properties, which could be due to substantial differences in network sizes. The relatively small correlation may suggest that players who are experienced in trade have a tendency to \emph{not} act out negative sentiments by spending money on bounties.
\end{itemize}

The exact values of the two correlation measures have to be interpreted with some caution. High values might be biased by e.g. the time a player spent in the game or by ignoring link weights for the number of exchanged private messages or traded money. Nevertheless, low values of $\rho(k_{\alpha},k_{\beta})$ indicate that hubs in one network are not necessarily hubs in another (see e.g. the Trade--Enmity case), suggesting that agents play very different roles in different relational networks. For example agents can be central for flows of information but peripheral for flows of goods \cite{roles}. In the \cite{SI} Text we give further relations between above network-network measures and study their evolutions in time (see Figs.~S1 and S2).

\subsection{Large-Scale Empirical Test of Structural Balance}
In the following we assign +(-)1 to a positively (negatively) connoted link. All friendship links have a value of +1, all enemy links -1. 
Social balance focuses on signed triads where the sign of a triad is the product of the signs of its three links.

 Social balance theory -- in its strong form \cite{cartwright} -- claims that positive triads are `balanced' while negative triads are `unbalanced', see Fig.~4.  Unbalanced triads are sources of stress and therefore tend to be avoided by agents when they update their personal relationships. From a physics point of view, the resulting dynamics can be viewed as an energy minimization process which may lead to jammed states \cite{antal} due to a rugged energy landscape \cite{marvel}. There is a  `weak formulation' of structural balance  \cite{davis} which postulates that triads with exactly two positive links are underrepresented in real networks, while the three other kinds of triads should be much more abundant. In the weak formulation only situations where ``the friend of my friend is my enemy'' are unstable, whereas in the strong form of structural balance, ``the enemy of my enemy is my enemy'' is also unstable, see Fig.~4.

 To test social balance we focus on the multiplex network of friendship and enmity interactions. The number of different types of triads are labeled $N_{\Delta}$. They are compared to the expected number of such triads in a null model (re-shuffled signs of links, $N_{\Delta}^{\rm rand}$, see \cite{SI}). In Fig.~4, a standard measure of statistical deviation, the $z$-score (see \cite{SI}), shows that $+++$ and $+--$ triads are heavily over-represented, while $++-$ triads are heavily under-represented with respect to pure chance. Triads of type $---$ are under-represented to a lesser degree than the three other types, favoring the weak formulation of structural balance over Heider's original formulation of balance theory. It is obvious that triads are characterized by different levels of stability. The robustness of these results is further confirmed by examining the time evolution of the number of triads in friendship/enmity networks over all 445 days, Fig.~5.
 
  \begin{figure}[b]
\includegraphics[width=0.45\textwidth]{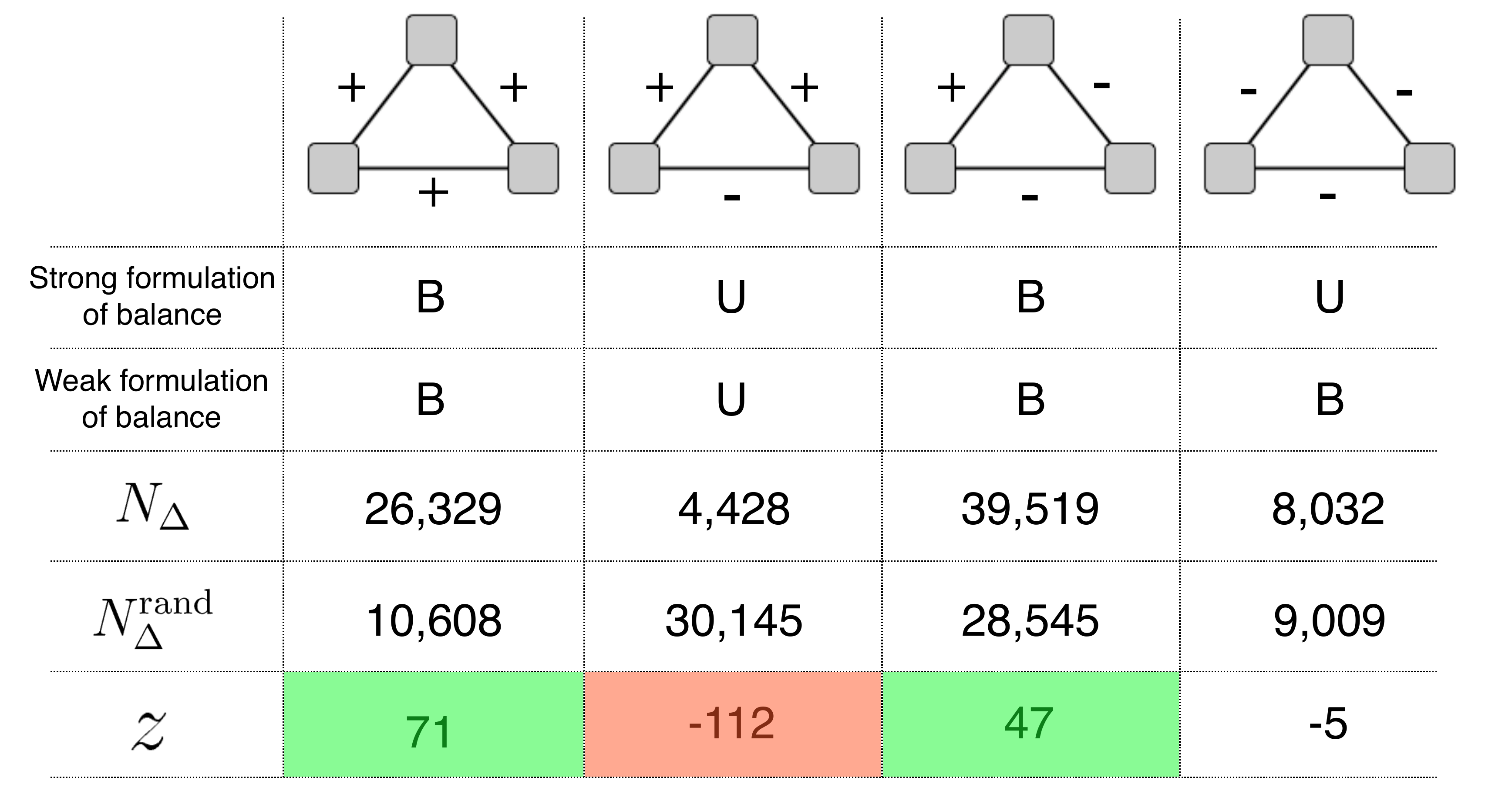}
\caption{Different types of signed triads, balanced (B) or unbalanced (U) according to the strong or weak formulation of structural balance. We show the number of each type of triad $N_{\Delta}$ in the friendship-enmity multiplex network, the expected number $N_{\Delta}^{\mathrm{rand}}$ of such triads when averaged over 1000 sign-randomizations and the corresponding $z$-score (see \cite{SI}). Triads $+++$ and $+--$ are over-represented, $++-$ triads are underrepresented with extraordinary significance.}\label{fig4}
\end{figure}

A detailed dynamical analysis of our data further reveals that a vast majority of changes in the network are due to the creation of new positive and negative links, not due to switching of existing links from plus to minus or vice versa. We illustrate this dominance of link destruction and creation over sign switching on the dynamics of the following triadic structures. Let us define a \emph{wedge} as a signed undirected triad with two links, i.e. a triad with one link missing (a `hole'). There are three possible wedge types: $++$, $+-$, $--$. We measure day-to-day transitions from wedges to other possible triadic structures. In the vast majority of all cases ($>99.9\%$), a wedge stays unchanged. In case of change, most often a hole is closed by either a positive or a negative link, see \cite{SI}~Fig.~3. The removal of a link is less frequent; sign switches almost never occur. This result is in marked contrast with many dynamical models of structural balance \cite{antal} which assume that a given social network is fully connected from the start and that only the link-signs are the relevant dynamical parameters, which evolve to reduce stress in the system. Our observation underpins that network sparsity and growth are fundamental properties and they need to be incorporated in any reasonable model of dynamics of positive and antagonistic forces in social systems. In full agreement with the results shown in Fig.~4 and Fig.~5, wedges of type $++$ close preferentially (about 7 times more likely) with a positive link, wedges of type $+-$ close preferentially (about 11 times more likely) with a negative link. There is no clear sign preference in the closure of type $--$ wedges. For details see \cite{SI} and \cite{SI}~Figs.~3--5.

We collect empirical transition rates in a transition matrix $M_{\mathrm{STC}}$, which we use in a simple dynamical model for \emph{Signed Triadic Closure (STC)}, see \cite{SI}. This STC model applies $M_{\mathrm{STC}}$ on a daily state vector of signed wedges. These wedges are closed or left unchanged according to the elements of $M_{\mathrm{STC}}$. With this model we are able to reproduce the empirical observations to a reasonable extent, see Fig.~5~(right).

 \begin{figure}[t]
\includegraphics{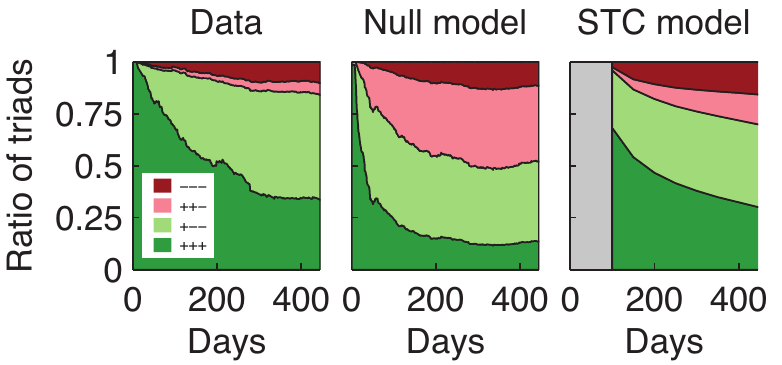}
\caption{Ratio of signed triad types over time. Left: Measured in the data. Center: Expected in the random null model, see \cite{SI}. Right: Simulation of signed triadic closure (STC) with a model based on wedge transition rates, see \cite{SI}. Initial condition: Measured network of day 100. All ratios measured in the data deviate significantly from ratios in the null model, except for the $---$ triads. The STC model reproduces the observed ratios considerably better.}\label{fig5}
\end{figure}

\section{Discussion}

Most empirical studies of large-scale social networks focus on node properties \cite{EL09}, for instance to uncover the topological centrality of social agents or patterns of homophily between agents \cite{obesity}, while being blind to the multiple nature of the links connecting agents. In many social systems, however, a proper description of multiplexity is essential to capture the stress caused by different forces acting on social agents and therefore to uncover the principles shaping the large-scale organization of social interactions. For instance, the interaction and co-existence of multiple relations are crucial to describe the emergence of conflict in social systems  \cite{conflict1,conflict2,conflict3} or the development of trust in commercial networks \cite{trust}.

Our work begins to quantitatively measure the multi-dimensionality of human relationships. Its results shed light on macroscopic implications of interaction types: Relations driven by aggression lead to markedly different systemic characteristics than relations of non-aggressive nature. Network-network interactions reveal a non-trivial structure of this multi-dimensionality, and how humans play very different roles in different relational networks. The richness of the data-set allows to explore the effect of multiple relations on the structure and stability of a large-scale social network, thereby providing a first empirical basis for the modeling of multiplex complex networks. Future research perspectives include different generalizations of structural balance theory, e.g. to a larger set of social relations, to the case of weighted and/or directed networks or to larger motifs, an extension of the concept of modularity for multiplex \cite{multiplex} or signed \cite{signed} networks but also dynamical aspects, for instance the dynamics of noncooperative organizations \cite{military}.

\section{Materials and Methods}
\subsection{Social Network Data from the Online Game `Pardus'}

The data-set contains practically all actions of all players of the MMOG Pardus (www.pardus.at) since 2004 when the game went online \cite{szell}. Pardus is an open-ended game with a world-wide player-base of more than 300,000 people. Players live in a virtual, futuristic universe which they explore and where they  interact  with others in a multitude of  ways to achieve their own goals \cite{castronova2005swb}. Here we focus on one of the three separate game {\em universes}, Artemis, in which $N=18,819$ players have interacted with at least one other player over the first 445 consecutive days of this universe's existence.

Players typically engage in various economic activities to accumulate wealth.  Communication between any two players can take place directly, by using a one-to-one, email-like , private message system  (PM), see \cite{SI}, or indirectly, by meeting in built-in chat channels or online forums. Social and economical decisions of players are often strongly influenced and driven by social factors such as friendship, cooperation and conflict. Conflictual relations may result in aggressive acts such as attacks, fights, revenge, even destruction of another player's means of production or transportation. Under certain conditions, hostile acts may degenerate into large-scale conflicts between different factions of players -- wars.

The Pardus data-set contains longitudinal and relational data allowing for an almost complete and dynamical mapping of multiplex relations  of an entire society. 
The data is free of interviewer effects since agents are not conscious of their actions being logged. Measurement errors which usually affect reliability of survey data \cite{carrington2005mam}, are practically absent. The longitudinal aspect of the data allows for the analysis of dynamical aspects such as the emergence  and evolution of network structures. Finally, it is possible to extract multiple social relationships between a fixed set of humans. We focus on the following set of six types of one-to-one interactions between players (for details see \cite{SI}): friendship and enmity relations, private message (PM) communication,  trades, attacks,  and revenge/punishment  through head money (bounties). 
We label these networks by Greek indices: $\alpha=1$ refers to  friendship networks, \ldots,   $\alpha=6$ to bounties. We focus on one-to-one interactions only (without projections as e.g. used in \cite{facebook,projectionBipartite1}) and discard indirect interactions such as mere participation in a chat.

Friendship and enmity networks are taken as snapshots at the last available day 445. All other networks are aggregated over time, meaning that whenever a link existed within day 1 and 445, it is counted as a link. For simplicity, we use unweighted, directed networks. Further, we define undirected networks as follows: A link exists between nodes $i$ and $j$ if there exists at least one directional link between those nodes. We construct triads (motifs of three connected nodes \cite{wasserman}) from undirected links.
For a combined analysis of the whole system we define an {\em envelope network} which is composed of the set of all links of \emph{all} interaction types. In the envelope network, a link from $i$ to $j$ exists if it exists in at least one of the six relational networks.

\subsection{Network Measures}
Network Measures. The statistical properties of the six networks have beenstudied as separate entities using the following notations and measures. $N_{\alpha}$ is the number of nodes in the network type $\alpha$, and $L_{\alpha}^{\mathrm{dir(undir)}}$ is the number of (un)directed links. Reciprocity is labeled by $r_{\alpha}$, and $\rho(k_{\alpha}^{\mathrm{in}},k_{\alpha}^{\mathrm{out}})$ is the correlation of in- and out-degrees within the $\alpha$ network. Average degree, clustering coefficient, and clustering coefficient with respect to the corresponding random
graph are marked by $\bar k_{\alpha}$, $C_{\alpha}$ and $C_{\alpha}/C_{\alpha}^{\mathrm{rand}}$, respectively. For more details, see the \cite{SI} Text.

\subsection{Network Interactions}
For network-network interactions, we compute the Jaccard coefficient which measures the interaction between two networks by measuring the tendency that links simultaneously are present in both networks. $J_{\alpha\beta}$ is a similarity score between two sets of elements and is defined as the size of the intersection of the sets divided by the size of their union \cite{jaccard}, $J_{\alpha\beta}\equiv |\alpha \cap \beta|/|\alpha \cup \beta|$. Related similarity measures, such as the cosine similarity measure lead to comparable results. The correlation measures used are described in detail in the \cite{SI}.

The legal department of the Medical University of Vienna has attested the innocuousness of the used anonymized data.

\subsection{ACKNOWLEDGEMENTS}
\begin{acknowledgments}
We thank Roberta Sinatra for helpful remarks. R.L. acknowledges support from the UK Engineering and Physical Sciences Research Council. This work was conducted within the framework of European Cooperation in Science and Technology Action MP0801 Physics of Competition and Conflicts. M.S. and S.T. acknowledge support from the Austrian Science Fund Fonds zur Förderung der wissenschaftlichen Forschung P 19132. During the redaction of this paper, we discovered an independent paper where the authors perform a large-scale verification of structural balance and arrive at conclusions similar to ours \cite{leskovec}.  \end{acknowledgments}

\end{article}

\end{document}